# Rough Solutions of the Einstein Constraint Equations on Compact Manifolds

**David Maxwell**

University of Alaska Fairbanks
E-mail: david.maxwell@uaf.edu

June 15, 2004

**Abstract.** We construct low regularity solutions of the vacuum Einstein constraint equations on compact manifolds. On 3-manifolds we obtain solutions with metrics in $H^s$ where $s > 3/2$. The constant mean curvature (CMC) conformal method leads to a construction of all CMC initial data with this level of regularity. These results extend a construction from [Ma04] that treated the asymptotically Euclidean case.

## 1. Introduction

Initial data for the vacuum Cauchy problem in general relativity consists of a Riemannian manifold $(M^3, g)$ and a symmetric $(0, 2)$-tensor $K$. One seeks an isometric embedding of $(M, g)$ into a Ricci-flat Lorenzian manifold such that $K$ is the extrinsic curvature of this embedding. Since the ambient manifold is Ricci flat, the initial data $(M, g, K)$ must satisfy the compatibility conditions

$$R_g - |K|_g^2 + \operatorname{tr}_g K^2 = 0$$
$$\operatorname{div} K - d \operatorname{tr}_g K = 0, \tag{1}$$

where $R_g$ is the scalar curvature of the metric $g$. System (1) is known known as the Einstein constraint equations; the first equation is the Hamiltonian constraint and the second is the momentum constraint.

The Cauchy problem has a formulation as a second-order quasilinear system of hyperbolic PDEs called the Einstein equations. In [CB52] Choquet-Bruhat proved that this problem is locally well-posed for sufficiently smooth initial data satisfying the constraint equations. Subsequent investigations have reduced the amount of regularity required from the initial data, but development of the low regularity theory of the constraint equations has lagged behind that of the evolution problem. For example, the local well-posedness theorem of [HKM77] for the Einstein equations permits initial data $(g, K)$ in $H^s_{\text{loc}} \times H^{s-1}_{\text{loc}}$ with $s > 5/2$. But solutions of the constraint equations with this level of regularity were not known to exist until recently [CBIY00]. The need for a theory of rough solutions of the constraint equations became especially prominent after the appearance of [KR] which proved an a priori estimate for the the time of existence of solutions of the vacuum Einstein equations depending on the size of $(\partial g, K)$ in $H^{s-1} \times H^{s-1}$ with $s > 2$.

Constructions of rough solutions of the constraint equations have been obtained in [CB04] on compact manifolds and in [CBIY00] and more generally in [Ma03] on asymptotically Euclidean manifolds. These works provide solutions with metrics in $H^2$. More specifically, these solutions were constructed in $W_{\text{loc}}^{k,p}$ spaces where $k \geq 2$ is an integer and $k > 3/p$. The restriction $k > 3/p$ is interesting because it is analogous to condition $s > 3/2$ for $H^s$ spaces. Since $s = 3/2$ is the scaling limit for the Einstein equations, this level of regularity is a natural lower bound for for seeking solutions of the constraint equations. These considerations motivated work in [Ma04] which gave a construction of all constant mean curvature (CMC) asymptotically Euclidean solutions of the constraint equations with metrics in $H_{\text{loc}}^s$ for real values of $s > 3/2$. In this paper we extend the low regularity construction from [Ma04] to compact manifolds. In particular, we show that the conformal method can be used to construct all CMC vacuum solutions $(g, K)$ of the constraint equations belonging to $H^s \times H^{s-1}$ with $s > 3/2$.

The constant mean curvature conformal method of Lichnerowicz [Li44] and Choquet-Bruhat and York [CBY80] is the basis of all known constructions of rough solutions of the constraint equations. The method starts with conformal data $(M^n, g, S, \tau)$ where $M^n$ is a smooth, compact, connected manifold with $n \geq 3$, $g$ is a metric on $M$, $S$ is a trace-free $(0, 2)$-tensor, and $\tau$ is a scalar function.[1] We then seek a solution $(\tilde{g}, \tilde{K})$ of the constraint equations on $M$ of the form

$$\tilde{g} = \phi^{2\kappa} g$$
$$\tilde{K} = \phi^{-2} \left( \sigma + \mathbb{L} W \right) + \frac{\tau}{n} \tilde{g}$$

where the unknowns $(\phi, W)$ are a positive function and vector field respectively, $\kappa = \frac{2}{n-2}$ is a dimensional constant, and $\mathbb{L}$ is the conformal Killing operator defined by

$$\mathbb{L} W_{ab} = \nabla_a W_b + \nabla_b W_a - \frac{2}{n} \operatorname{div} W g_{ab}.$$

It follows that $(\tilde{g}, \tilde{K})$ solves the constraint equations so long as

$$-a\Delta_g \phi + R_g \phi = |S + \mathbb{L} W|_g^2 \phi^{-2\kappa-3} - b\tau^2 \phi^{2\kappa+1}$$
$$\operatorname{div} \mathbb{L} W = b\phi^{-n\kappa} \, d\tau - \operatorname{div} S \tag{2}$$

where $R_g$ is the scalar curvature of $g$ and $a = 2\kappa + 4$ and $b = \frac{n-1}{n}$ are constants. The function $\tau$ specifies the mean curvature $\operatorname{tr}_{\tilde{g}} \tilde{K}$, and the conformal method is most successful when $\tau$ is assumed to be constant (see, however, [MI96][CB04][IOM04] for results in the non-CMC case). When $\tau$ is constant, system (2) decouples and the second equation becomes

$$\operatorname{div} \mathbb{L} W = -\operatorname{div} S \tag{3}$$

---

[1] We work with general $n$-dimensional manifolds since some of our auxiliary results are meaningful outside the context of the constraint equations.

In the smooth setting, equation (3) is always solvable and leads to a decomposition of the set of smooth symmetric, trace-free $(0, 2)$-tensors $\mathscr{S}$

$$\mathscr{S} = \operatorname{im} \mathbb{L} \oplus \ker \operatorname{div} \tag{4}$$

known as York splitting. Tensors in $\mathscr{S} \cap \ker \operatorname{div}$ are known as transverse-traceless, and specifying conformal data $(M, g, S, \tau)$ for the CMC conformal method is equivalent to specifying $(M, g, \sigma, \tau)$, where $\sigma$ is the transverse-traceless part of $S$. The study of system (2) reduces to determining conditions when the Lichnerowicz equation

$$-a\Delta_g \phi + R_g \phi = |\sigma|_g^2 \phi^{-2\kappa-3} - b\tau^2 \phi^{2\kappa+1} \tag{5}$$

has a positive solution. For smooth metrics it is known[CBY80][Is95] that (5) is solvable if and only if one of the following conditions hold:

1. $\sigma \not\equiv 0$ and $\tau \neq 0$,

2. $\lambda_g > 0$, $\sigma \not\equiv 0$ and $\tau = 0$,

3. $\lambda_g = 0$, $\sigma \equiv 0$ and $\tau = 0$,

4. $\lambda_g < 0$, $\sigma \equiv 0$ and $\tau \neq 0$.

Here $\lambda_g$ is the Yamabe invariant defined by

$$\lambda_g = \inf_{\substack{f \in C^\infty(M) \\ f \neq 0}} \frac{\int_M a\, |\nabla f|_g^2 + R_g f^2 \, dV_g}{||f||_{L^{2^*}}^2}.$$

Our main result, Theorem 4.5, shows that these conditions are also necessary and sufficient for the solvability of (5) when $g \in H^s$ and $\sigma \in H^{s-1}$ with $s > n/2$. We also show in Theorem 5.2 that the York decomposition (4) holds for symmetric, trace-free $(0, 2)$-tensors in $H^{s-1}$ and hence the CMC conformal method fully extends to the rough setting.

Aside from lowering the regularity of solutions, we also obtain some technical improvements to the previous construction of rough solutions on compact manifolds obtained in [CB04]. Solutions were found in that reference in the Yamabe negative case ($\lambda_g < 0$) only under an additional hypothesis on the singularities of $\sigma$, namely $\sigma \in H^{s-1} \cap L^\infty$. Also, the case $\lambda_g > 0$, $\tau = 0$ and $\sigma \not\equiv 0$ when $\sigma$ has zeros was not treated. The zeros and singularities of $\sigma$ pose difficulties for the method of sub- and supersolutions used to find solutions of (5). Our approach to finding sub- and supersolutions of (5), discussed in Section 4, circumvents these problems by means of an initial conformal transformation to a metric with the singularities and zeros of $\sigma$ already encoded in the scalar curvature. This strategy has been used previously for smooth conformal data, but the choice of conformal change used here is new and has interest even in the context of smooth solutions.

## 2. Considerations for Rough Metrics

Before turning to the specifics of solving the constraint equations, we start with an outline of the rules for working with rough metrics. Throughout this paper, let $(M^n, g)$ be a compact, connected Riemannian manifold with $n \geq 3$ and $g \in H^s$. Our definitions and notation for Sobolev spaces follows [Ta96]. The value $s = n/2$ is a dividing point between metrics that are natural for geometric analysis and those that are not. For example, the Laplacian of a metric in $H^s$ with $s < n/2$ doesn't necessarily take $H^2$ to $L^2$ or even $H^1$ to $H^{-1}$. This is a consequence of the fact that a metric in $H^s$ with $s < n/2$ is not necessarily in $L^\infty$. On the other hand, a metric in $H^s$ with $s > n/2$ can be nearly as good as a smooth metric. The following well-known lemma is of central importance.

**Lemma 2.1.** *Suppose $M^n$ is a compact manifold, $\sigma \leq \min(s_1, s_2)$, $s_1 + s_2 \geq 0$ and $\sigma < s_1 + s_2 - \frac{n}{2}$. Then pointwise multiplication extends to a continuous bilinear map*

$$H^{s_1}(M) \times H^{s_2}(M) \to H^\sigma(M).$$

Lemma 2.1 implies that if $u \in H^s$ with $s > n/2$, then multiplication by $u$ takes $H^\sigma$ to $H^\sigma$ for every $\sigma \in [-s, s]$. In particular, $H^s$ is an algebra. This fact allows us to have well defined conformal changes of metric, for example. If $\phi$ is a positive function in $H^s$, and if $g$ is a metric in $H^s$, then Lemma 2.1 implies $\phi^\alpha g$ is again rough metric in $H^s$ for any positive integer $\alpha$. To allow arbitrary real exponents $\alpha$, and to work with other nonlinearities, we require the following consequence of Lemma 2.1.

**Lemma 2.2.** *Suppose $M^n$ is a compact manifold and $F : I \to \mathbb{R}$ is smooth on some interval $I$. If $s > n/2$ and $u \in H^s(M)$ satisfies $u(x) \in I$ for all $x \in M$, then $F(u) \in H^s(M)$.*

*Proof:* Working in a coordinate chart we can assume that $u$ is defined on $\mathbb{R}^n$ and need only show that $F(u) \in H^s_{\text{loc}}$. Let $k$ be the largest integer such that $k \leq s$. If $\alpha$ is a multi-index such that $|\alpha| = k$, then $\partial_\alpha F(u)$ is a sum of terms of the form $F^{(r)}(u)\partial_{\alpha_1} u \cdots \partial_{\alpha_r} u$ where $\Sigma_{i=1}^r |\alpha_i| = k$. From repeated applications of Lemma 2.1 we have $\partial_{\alpha_1} u \cdots \partial_{\alpha_r} u \in H^\sigma_{\text{loc}}$ for any $\sigma < s - k + (r-1)(s - \frac{n}{2})$. Since $F^r(u) \in L^\infty_{\text{loc}}$, this shows that $F(u) \in H^k_{\text{loc}}$. To obtain the full regularity of $F(u)$, consider any derivative $\partial_i F(u) = F'(u)\partial_i u$. Since $F'$ is smooth, we have $F'(u) \in H^k_{\text{loc}}$ and from Lemma 2.1 we have $F'(u)\partial_i u \in H^{\sigma_1 - 1}$ for every $\sigma_1 \leq s$ such that $\sigma_1 < k + s - \frac{n}{2}$. Hence $F(u) \in H^{\sigma_1}_{\text{loc}}$. Repeating this argument shows that $F(u) \in H^{\sigma_2}_{\text{loc}}$ for every $\sigma_2 \leq s$ such that $\sigma_2 < k + 2\left(s - \frac{n}{2}\right)$. A finite number of further repetitions shows that $F(u) \in H^s_{\text{loc}}$.  $\square$

Consider the Laplacian $\Delta_g$ of a rough metric $g$. In local coordinates we have

$$\Delta_g u = \frac{1}{\sqrt{\det g}} \partial_i (\sqrt{\det g} g^{ij} \partial_j u).$$

It follows from Lemmas 2.1 and 2.2 that the Laplacian has the form in local coordinates

$$a^{ij}\partial_i\partial_j + b^i\partial_i$$

where $a \in H^s_{\text{loc}}$ and $b^i \in H^{s-1}_{\text{loc}}$. Using Lemma 2.1 it follows that $\Delta_g : H^\sigma \to H^{\sigma-2}$ for any $\sigma \in [2 - s, s]$. The following class of differential operators encodes the regularity of the coefficients of the Laplacian.

**Definition 2.3** *Let $A$ be a linear differential operator that in any coordinate chart has the form*

$$A = \sum_{|\alpha|\le m} a^\alpha \partial_\alpha,$$

*where $a^\alpha$ is a $\mathbb{R}^{k\times k}$ valued function. We say that*

$$A \in \mathscr{L}^{m,s}$$

*if $a^\alpha \in H^{s-m+|\alpha|}_{\text{loc}}$ for all $|\alpha| \le m$.*

An analogous definition holds for differential operators on sections of vector bundles. From Lemma 2.1 we have the following mapping properties.

**Lemma 2.4.** *Suppose $m \le 2s$ and $\sigma \in [m - s, s]$. If $A \in \mathscr{L}^{m,s}$, then $A$ defines a continuous map*

$$A : H^\sigma \to H^{\sigma-m}.$$

We say $A$ is elliptic if for each $x$, the constant coefficient operator $\Sigma_{|\alpha|=m} a^\alpha(x)\partial_\alpha$ is elliptic in the usual sense. Using Lemmas 2.1 and 2.2 it is easy to check that the Laplacian $\Delta_g$, conformal Laplacian $-a\Delta_g + R_g$, and vector Laplacian $\Delta_{\mathbb{L}} = \text{div}\,\mathbb{L}$ all belong to $\mathscr{L}^{2,s}$ when $s > n/2$ and are all elliptic.

The following interior elliptic estimate was proved in [Ma04]. We use the notation $A \lesssim B$ to mean $A < cB$ for a positive constant $c$ where $c$ does not depend on the variable functions and parameters of $A$ and $B$. For example, in estimate (6) below the implicit constant is independent of the function $u$ but can can depend on the open sets $U$, $V$, the operator $A$, and so forth.

**Proposition 2.5.** *Let $U$ and $V$ be open sets in $\mathbb{R}^n$ with $U \subset\subset V$, and suppose $s > n/2$ and $\sigma \in (m - s, s]$. If $A \in \mathscr{L}^{m,s}$ is elliptic, then for every $u \in H^\sigma$ we have*

$$||u||_{H^\sigma(U)} \lesssim ||Au||_{H^{\sigma-m}(V)} + ||u||_{H^{m-s}(V)}. \tag{6}$$

Using coordinate charts and cutoff functions, we obtain as an immediate consequence a-priori estimates on compact manifolds.

**Corollary 2.6.** *Suppose* $(M^n, g)$ *is a compact Riemannian manifold with* $g \in H^s$ *and* $s > n/2$. *If* $A$ *is a second order elliptic differential operator in* $\mathcal{L}^{2,s}$, *then*

$$||X||_{H^s(M)} \lesssim ||AX||_{H^{s-2}(M)} + ||X||_{L^2(M)}$$

*for all* $X \in H^s(M)$.

We now turn to duality properties and integration by parts for rough metrics. We would like to claim, for example, that the equation

$$\int_M (-a\Delta_g u + R_g u)v \; dV_g = \int_M a \, |\nabla u|_g^2 + R_g uv \; dV_g$$

that holds for the conformal Laplacian of a smooth metric also holds for rough metrics. We need to show that integration by parts holds and that the term $\int_M R_g uv \; dV_g$ can be suitably interpreted for rough metrics (note that $R_g$ need not be a locally integrable function when $n = 3$).

Suppose $g$ is a rough metric in $H^s$ with $s > n/2$ and $\hat{g}$ is a smooth background metric on $M$. We have a bilinear form defined for $u, v \in C^\infty(M)$ by

$$\langle u, v \rangle_{(M, \hat{g})} = \int_M uv \; dV_{\hat{g}} \tag{7}$$

which extends for any $\sigma \in \mathbb{R}$ to a continuous bilinear map

$$\langle \cdot, \cdot \rangle_{(M, \hat{g})} : H^\sigma \times H^{-\sigma}(M) \to \mathbb{R}$$

and induces an isomorphism

$$H^\sigma(M) = \left(H^{-\sigma}(M)\right)^*.$$

In applications we would prefer use the rough metric $g$ instead of the smooth metric $\hat{g}$ to induce this isomorphism. This can be done for a restricted range of Sobolev spaces. From Lemma 2.2, we have $dV_g = h \; dV_{\hat{g}}$ where $h > 0$ and $h \in H^s(M)$. Multiplication by $h$ is continuous on on $H^\sigma$ for every $\sigma \in [-s, s]$, so we can define for $u \in H^\sigma$ and $v \in H^{-\sigma}$

$$\langle u, v \rangle_{(M, g)} = \langle hu, v \rangle_{(M, \hat{g})}.$$

With this definition, we have the following easy consequence.

**Lemma 2.7.** *Suppose* $(M^n, g)$ *is a compact Riemannian manifold with* $g \in H^s$ *and* $s > n/2$ *and suppose* $\sigma \in [-s, s]$. *Then for every* $u \in H^\sigma$ *and* $v \in H^{-\sigma}$,

$$\left|\langle u, v \rangle_{(M, g)}\right| \lesssim ||u||_{H^\sigma} ||v||_{H^{-\sigma}}.$$

*Moreover, $H^\sigma(M)$ is naturally isomorphic to $\left(H^{-\sigma}(M)\right)^*$ through the pairing $\langle \cdot, \cdot \rangle_{(M,g)}$. Finally, for smooth functions $u$ and $v$,*

$$\langle u, v \rangle_{(M,g)} = \int_M uv \ dV_g.$$

These ideas readily extend to sections of vector bundles as well. We show here explicitly how this is done for sections of the tangent bundle. The smooth background metric $\hat{g}$ gives rise to a bilinear form defined for smooth vector fields $X$ and $Y$ by

$$\langle X, Y \rangle_{(M,\hat{g})} = \int_M \langle X, Y \rangle_g \ dV_{\hat{g}}.$$

This map extends to a continuous bilinear form of less regular vector fields

$$\langle \cdot, \cdot \rangle_{(M,\hat{g})} : H^\sigma \times H^{-\sigma} \to \mathbb{R}$$

for every $\sigma \in \mathbb{R}$. Now if $X$ is a $H^\sigma$ vector field with $\sigma \in [-s, s]$, we can define $X_* \in H^\sigma$ by

$$X_*^a = h \hat{g}^{ab} g_{bc} X^c$$

where as before $dV_g = h \ dV_{\hat{g}}$. We then define for vector fields $X$ and $Y$ in $H^\sigma$ and $H^{-\sigma}$ respectively

$$\langle X, Y \rangle_{(M,g)} = \langle X_*, Y \rangle_{(M,\hat{g})},$$

and we note that if $X$ and $Y$ are smooth then

$$\langle X, Y \rangle_{(M,g)} = \int_M \langle X, Y \rangle_g \ dV_g.$$

To establish an integration by parts formula, suppose that $u \in H^\sigma$ with $\sigma \in [2-s, s]$, $v \in H^{2-\sigma}$, and suppose that $v$ is supported in the domain of some coordinate chart. We then have in local coordinates

$$\langle \nabla u, \nabla v \rangle_{(M,g)} = \sum_{i,j} \left\langle \sqrt{\det g} \, g^{ij} \partial_i u, \partial_j v \right\rangle_{(\mathbb{R}^n, \overline{g})}$$

where the bilinear form on the right hand side is the usual pairing of $H^{\sigma-1}$ against $H^{1-\sigma}$ for the flat Euclidean metric. The standard integration by parts formula

$$\langle \partial_i f, g \rangle_{(\mathbb{R}^n, \overline{g})} = \langle f, -\partial_i g \rangle_{(\mathbb{R}^n, \overline{g})}$$

is valid for $f \in H^\tau(\mathbb{R}^n)$ and $g$ compactly supported in $H^{1-\tau}(\mathbb{R}^n)$, for any $\tau \in \mathbb{R}$. Hence

$$\sum_{i,j} \left\langle \sqrt{\det g} \, g^{ij} \partial_i u, \partial_j v \right\rangle_{\mathbb{R}^n, \overline{g}} = \sum_{i,j} \left\langle -\partial_j \sqrt{\det g} \, g^{ij} \partial_i u, v \right\rangle_{\mathbb{R}^n, \overline{g}} = \langle -\Delta_g u, v \rangle_{(M,g)}.$$

A partition of unity argument then establishes

$$\langle \nabla u, \nabla v \rangle_{(M,g)} = \langle -\Delta_g u, v \rangle_{(M,g)}$$

for every $u \in H^\sigma(M)$ and $v \in H^{2-\sigma}(M)$ with $\sigma \in [2-s, s]$. A similar approach also shows that integration by parts also holds for the vector Laplacian.

**Lemma 2.8.** *Suppose $(M^n, g)$ is a compact Riemannian manifold with $g \in H^s$ with $s > n/2$. Then for functions $u \in H^\sigma$ and $v \in H^{2-\sigma}$ with $\sigma \in [2-s, s]$ we have*

$$\langle \nabla u, \nabla v \rangle_{(M,g)} = \langle -\Delta_g u, v \rangle_{(M,g)}.$$

*For vector fields $X \in H^\sigma$ and $Y \in H^{2-\sigma}$ with $\sigma \in [2-s, s]$ we have*

$$\langle \mathbb{L}X, \mathbb{L}Y \rangle_{(M,g)} = \langle -\Delta_{\mathbb{L}} X, Y \rangle_{(M,g)}.$$

Finally, we show that maximum principles hold for rough metrics. These are used most importantly to ensure that the conformal factors we construct are positive. We say that a distribution $V \in H^\sigma$ with $\sigma \in [-s, s]$ satisfies $V \geq 0$ if

$$\langle V, \phi \rangle_{(M,g)} \geq 0$$

for every smooth function $\phi \geq 0$.

**Lemma 2.9.** *Suppose $(M^n, g)$ is a compact Riemannian manifold with $g \in H^s$ and $s > n/2$. Suppose also that $V \in H^{s-2}$ and that $V \geq 0$ with $V \not\equiv 0$. If $u \in H^s$ satisfies*

$$-\Delta u + V u \leq 0, \tag{8}$$

*then $u \leq 0$.*

*Proof:* Let $v = u^{(+)}$. One readily finds from Lemma 2.1, since $u^{(+)} \in H^1$, that $uv$ belongs to $H^1$. Since $s - 2 \geq -1$ and since $V \in H^{s-2}$, we can apply $V$ to $uu^{(+)}$. Since $V \geq 0$, and since $uv \geq 0$, we have $\langle V, uv \rangle_{(M,g)} \geq 0$. Finally, since $u$ satisfies (8) we obtain

$$\langle -\Delta u, v \rangle_{(M,g)} \leq - \langle V, uv \rangle_{(M,g)} \leq 0.$$

Integrating by parts and using the fact that

$$\nabla v = \{ \; x \quad y = 1$$

we conclude $v$ is constant. To show that $v \equiv 0$ we see that if not, then $u^{(+)} = v$ is a positive constant and hence $u$ is a positive constant. Since $-\Delta u + V u = V u \geq 0$, and since $-\Delta u + V u \leq 0$ we conclude $V \leq 0$ and $V \geq 0$. But then $V \equiv 0$, a contradiction. $\square$

A version of the following strong maximum principle was proved in [Ma04] for asymptotically Euclidean metrics. Since the argument there was purely local, using only the regularity of the metric and the connectivity of the manifold, the same proof yields a strong maximum principle for compact manifolds.

**Proposition 2.10.** *Suppose $(M^n, g)$ is a compact Riemannian manifold with $g \in H^s$ and $s > n/2$. Suppose that $u \in H^s(M)$ is nonnegative and that $V \in H^{s-2}(M)$ satisfies*

$$-\Delta_g u + V u \geq 0.$$

*If $u(x) = 0$ at some point $x \in M$, then $u$ vanishes identically.*

# 3. The Yamabe Invariant of a Rough Metric

The Yamabe invariant for smooth compact manifolds, defined by

$$\lambda_g = \inf_{\substack{f \in C^\infty(M) \\ f \not\equiv 0}} \frac{\int_M a \left|\nabla f\right|_g^2 + R_g f^2 \; dV_g}{||f||^2_{L^{2^*}}},$$

where $2^* = 2n/(n-2)$, plays an essential role in classifying CMC solutions of the constraint equations on compact manifolds. To understand properties of the Yamabe invariant for rough metrics, we consider the quadratic bilinear form

$$A(f, h) = \int_M a \left\langle \nabla f, \nabla h \right\rangle + R_g f h \; dV_g.$$

When $g$ is smooth then $A$ is obviously continuous on $H^1 \times H^1$. This is also true when $g$ is rough, so long as we interpret $\int_M R_g f h \; dV_g$ as $\langle R_g, f h \rangle_{(M,g)}$. To see this, we note that the term $\int_M a \left\langle \nabla f, \nabla h \right\rangle \; dV_g$ poses no difficulties and we need only consider the scalar curvature term. From Lemmas 2.1 and 2.8 we have

$$\left| \langle R_g, f h \rangle_{(M,g)} \right| \lesssim ||R_g||_{H^{s-2}} ||f||_{H^1} ||h||_{H^1} \tag{9}$$

since $1 + 1 - \frac{n}{2} > 2 - s$. This establishes the continuity of $A$ on $H^1$. In fact, we can say more than this. Pick $\eta$ such that $0 < \eta < 1$ and such that $2\eta < s - \frac{n}{2}$. Then $(1 - \eta) + (1 - \eta) - \frac{n}{2} > 2 - s$ as well, and we conclude from Lemma 2.1 that

$$\left| \langle R_g, f h \rangle_{(M,g)} \right| \lesssim ||R_g||_{H^{s-2}} ||f||_{H^{1-\eta}} ||h||_{H^{1-\eta}}. \tag{10}$$

Hence the scalar curvature term is continuous on $H^{1-\eta} \times H^{1-\eta}$.

We would like to show that $\lambda_g \neq -\infty$, which we do by considering the first eigenvalue of the conformal Laplacian. Let

$$J_p(f) = \frac{A(f, f)}{||f||^2_{L^p}},$$

where $||f||_{L^p}$ is the metric dependent norm defined by

$$||f||^p_{L^p} = \int_M |f|^p \; dV_g.$$

We have

$$\lambda_g = \inf_{\substack{f \in C^\infty(M) \\ f \not\equiv 0}} J_{2^*}(f)$$

and the first eigenvalue of the conformal Laplacian is defined by

$$\mu_h = \inf_{\substack{f \in C^\infty(M) \\ f \not\equiv 0}} J_2(f).$$

If $\lambda_g < 0$ then it is easy to see that $\mu_g \leq \lambda_g$ and therefore $\lambda_g \neq -\infty$ if $\mu_g \neq -\infty$. Our estimate for a lower bound for $\mu_g$ starts with the following estimate for $A$.

**Lemma 3.1.** *Let $(M^n, g)$ be a compact Riemannian manifold with $g \in H^s$ and $s > n/2$. Then there are positive constants $C_1$ and $C_2$ such that for every $f \in H^1$*

$$A(f, f) \geq C_1 ||f||_{H^1}^2 - C_2 ||f||_{L^2}^2$$

*Proof:* Pick $\eta$ as was done in deriving inequality (10). From interpolation we have for any $\epsilon > 0$ that

$$||f||_{H^{1-\eta}} \leq C(\eta, \epsilon) ||f||_{L^2} + \epsilon ||f||_{H^1}.$$

From inequality (10) we then have

$$\left| \left\langle R_g, f^2 \right\rangle_{(M,g)} \right| \leq C(\eta, \epsilon, ||R_g||_{H^{s-2}}) ||f||_{L^2}^2 + \epsilon^2 ||f||_{H^1}^2. \tag{11}$$

We also note that

$$||f||_{H^1}^2 \lesssim \int_M a \, |\nabla f|_g^2 \, dV_g + ||f||_{L^2}^2. \tag{12}$$

Since

$$A(f, f) \geq \int_M a \, |\nabla f|_g^2 \, dV_g - \left| \left\langle R_g, f^2 \right\rangle_{(M,g)} \right|$$

we obtain from (11) and (12), taking $\epsilon$ sufficiently small, that

$$A(f, f) \gtrsim ||f||_{H^1}^2 - \left( C(\eta, \epsilon, ||R_g||_{H^{s-2}}) + 1 \right) ||f||_{L^2}^2.$$

□

It follows immediately that $\lambda_g$ and $\mu_g$ are both finite.

**Lemma 3.2.** *Let $(M^n, g)$ be a compact Riemannian manifold with $g \in H^s$ and $s > n/2$. Then $\lambda_g$ and $\mu_g$ are both finite.*

*Proof:* Since $||f||_{L^2}^2 \lesssim ||f||_{L^{2^*}}^2$, it follows that if $\lambda_g = -\infty$ then $\mu_g = -\infty$. So it is enough to show that $\mu_g$ is finite. From Lemma 3.1 we have

$$\int_M a \, |\nabla f|_g^2 + R_g f^2 \, dV_g \geq C_1 ||f||_{H^1}^2 - C_2 ||f||_{L^2}^2 \geq -C_2 ||f||_{L^2}^2.$$

Dividing by $||f||_{L^2}^2$ we conclude that

$$\mu_g \geq -C_2.$$

□

**Proposition 3.3.** *Let* $(M^n, g)$ *be a compact Riemannian manifold with* $g \in H^s$ *and* $s > n/2$. *Then the following are equivalent:*

1. *There exists a positive function* $\phi \in H^s$ *such that* $\tilde{g} = \phi^{2\kappa} g$ *satisfies* $R_{\tilde{g}}$ *is continuous and positive (resp. zero, resp. negative).*

2. $\lambda_g$ *is positive (resp. zero, resp. negative).*

3. $\mu_g$ *is positive (resp. zero, resp. negative).*

*Proof:* We start by proving Condition 3 implies Condition 1. We claim that there is function $f$ in $H^1$, not identically zero, such that $J_2(f) = \mu_g$. To see this, let $\{f_k\}$ be a sequence of smooth functions such that $||f_k||_{H^1} = 1$ and $J_2(f_k)$ converges to $\mu_g$. Since the sequence is bounded in $H^1$, we can assume without loss of generality that it converges weakly to some $f \in H^1$. From Lemma 3.1 we have

$$J_2(f_k) \geq \frac{C_1 ||f_k||_{H^1}^2 - C_2 ||f_k||_{L^2}^2}{||f_k||_{L^2}^2} = \frac{C_1}{||f_k||_{L^2}^2} - C_2,$$

Since $f_k \to f$ strongly in $L^2$, and since the sequence $J_2(f_k)$ is bounded, we conclude that $f$ is not identically zero. It is easy to see that $J_2$ is weakly lower semicontinuous on $H^1 \setminus \{0\}$ and hence we conclude that $J_2(f) = \mu_g$.

Since $f$ is a minimizer in $H^1$ of $J_2(f)$ it satisfies

$$-a\Delta_g f + R_g f = \mu_g f$$

in $H^{-1}$. Proposition 6.1 then implies $f \in H^s$. We can assume without loss of generality that $f \geq 0$ since $|f| \in H^1$ and $J_2(f) = J_2(|f|)$. So from the strong maximum principle Proposition 2.10 we conclude that $f > 0$ everywhere. Letting $\phi = f$ and $\tilde{g} = \phi^{2\kappa} g$, it follows that $R_{\tilde{g}} = \phi^{-2\kappa-1}(-a\Delta_g \phi + R_g \phi) = \phi^{-2\kappa} \mu_g$ Since $\phi$ is continuous and positive, we conclude that $R_{\tilde{g}}$ is continuous and everywhere has the same sign as $\mu_g$.

We next show Condition 1 implies Condition 2. Since $\lambda_g$ is a conformal invariant, we can assume without loss of generality that $R_g$ is continuous and has constant sign. If $R_g$ is negative, then $J_{2^*}(1) < 0$ and hence $\lambda_g < 0$. If $R_g \geq 0$ then clearly $\lambda_g \geq 0$. But if $R_g = 0$, then $J_{2^*}(1) = 0$ so $\lambda_g = 0$. If $R_g > 0$, then

$$J_{2^*}(f) \gtrsim ||f||_{H^1}^2 \gtrsim ||f||_{L^{2^*}}^2.$$

and hence $\lambda_g > 0$. On the other hand, if $R_g = 0$, then $J_{2^*}(1) = 0$ and hence $\lambda_g = 0$.

Finally, we show that Condition 2 implies Condition 3. If $\lambda_g < 0$, then $A(f, f) < 0$ for some smooth function $f$, so $\mu_g < 0$ also. On the other hand, if $\lambda_g \geq 0$, then $A(f, f) \geq 0$ for every smooth function $f$ and $\mu_g \geq 0$ also. Since $||f||_{L^2}^2 \lesssim ||f||_{L^6}^2$ we conclude that $J_{2^*}(f) \lesssim J_2(f)$ for every smooth non-vanishing function $f$. Hence if $\lambda_g > 0$, then $\mu_g > 0$. On the other hand, from previous implications we have shown that if $\mu_g > 0$, then $\lambda_g > 0$. So if $\lambda_g = 0$ then $\mu_g = 0$, which competes the proof.  $\square$

We conclude by showing that the sign of $\lambda_g$ can be detected if $(M, g)$ has scalar curvature with $R_g \geq 0$ or $R_g \leq 0$ even when $R_g$ is not continuous.

**Corollary 3.4.** *Suppose $(M, g)$ is a compact Riemannian manifold with $g \in H^s$ and $s > n/2$. If $R_g \leq 0$ then $\lambda_g \leq 0$ and $\lambda_g = 0$ if and only if $R_g = 0$. If $R_g \geq 0$ then $\lambda_g \geq 0$ and $\lambda_g = 0$ if and only if $R_g = 0$.*

*Proof:* Suppose $R_g \leq 0$. Computing $J_{2^*}(1)$ we find $\lambda_g \leq 0$. If $\lambda_g = 0$ then $\langle R_g, c \rangle_{(M,g)} = 0$ for any constant $c$. Since $R_g \leq 0$, it follows that for any smooth function $f$

$$0 = \left\langle R_g, \min_M f \right\rangle_{(M,g)} \geq \langle R_g, f \rangle_{(M,g)} \geq \left\langle R_g, \max_M f \right\rangle_{(M,g)} = 0$$

and from Lemma 2.7 we conclude $R_g = 0$.

If $R_g \geq 0$ then it is clear that $\lambda_g \geq 0$. If $R_g \geq 0$ and $\lambda_g = 0$ then $\mu_g = 0$ also. Let $\{f_n\}$ be a sequence of non-negative functions in $H^1$ such that $J_2(f_n) \to \mu_g = 0$ and such that $||f_n||_{L^2} = 1$. Then we have $||f_n||_{H^1}$ is bounded and some subsequence converges weakly in $H^1$ to a function $f$ such that $||f||_{L^2} = 1$. From the weak lower semicontinuity of $J_2$ we have $J_2(f) = 0$ and hence $\nabla f = 0$ almost everywhere. So $f$ is a constant and moreover $\langle R_g, f^2 \rangle = 0$. This implies $\langle R_g, c \rangle = 0$ for any constant $c$, and using the fact that $R_g \geq 0$ a similar argument as before implies $R_g = 0$.

Finally, we note that if $R_g = 0$, then Proposition 3.3 implies $\lambda_g = 0$.  $\square$

## 4. The Hamiltonian Constraint

In this section we determine conditions under which the Lichnerowicz equation

$$-a \Delta_g \phi + R_g \phi = |\sigma|_g^2 \phi^{-2\kappa-3} - b\tau^2 \phi^{\kappa+1} \tag{13}$$

is solvable. Although we only require solutions when $\tau$ is constant, it might be useful for future considerations of non-CMC rough solutions to treat here the case of a general function $\tau$. Hence we seek necessary and sufficient properties of the conformal data $(M, g, \sigma, \tau)$ leading to the existence of a positive solution of (13), including the case where $\tau$ is not constant.

A standard approach to finding a solution to (13) is to first make a conformal change to a metric with amenable scalar curvature and then to seek constant sub- and super solutions $\phi_-$ and $\phi_+$ such that $0 < \phi_- \le \phi_+$. Having done this, a barrier argument proves that there exists a solution $\phi$ with $0 < \phi_- \le \phi \le \phi_+$. Propositions 6.2 and 6.4 of Section 6 show that this same barrier argument works for rough metrics. So we concentrate here of finding sub- and supersolutions.

To illustrate the potential for difficulty in finding constant sub and supersolutions for rough conformal data, consider the case $\lambda_g > 0$, and let us suppose for simplicity that the scalar curvature $R_g \equiv 1$. We then seek constants $\phi_- \le \phi_+$ such that

$$\phi_+ \ge |\sigma|^2_g \, \phi_+^{-2\kappa-3} - b\tau\phi_+^{\kappa+1}$$
$$\phi_- \le |\sigma|^2_g \, \phi_-^{-2\kappa-3} - b\tau\phi_-^{\kappa+1}.$$

If the conformal data are smooth, $\tau$ is constant, and $\sigma$ is non-vanishing, then the existence of such constants follows easily from the asymptotic behaviour of the functions $y \mapsto y^{-2\kappa-3}$ and $y \mapsto -y^{\kappa+1}$. But if $\sigma$ vanishes somewhere we cannot find a positive constant such that

$$\phi_- < |\sigma|^2_g \, \phi_-^{-2\kappa-3} - b\tau\phi_-^{\kappa+1}$$

everywhere on $M$. Similarly, if $\sigma$ has singularities, then we typically cannot find a constant supersolution either. Since the difficulties posed by zeros are similar to those posed by singularities, we consider how the problem of zeros is treated for smooth conformal data.

Suppose that $(M, g)$ is smooth, $R_g \equiv 1$, $\tau \equiv 0$ and suppose $\sigma \not\equiv 0$ has zeros. A solution to the Lichnerowicz equation in this case was found in [Is95] by finding a non-constant subsolution. Specifically, $\phi_-$ can be taken to be the solution of

$$-a\Delta_g\phi_- = |\sigma|^2_g \, A$$

where $A = \max\{1, \max_M |\sigma|^2_g\}$. An argument using the monotonicity of the function $y \mapsto y^{-2\kappa-3}$ and the maximum principle then shows that $\phi_-$ is a subsolution. Although this argument cannot be adapted directly to the situation where $\sigma$ might have singularities, there is another way to understand the role that this subsolution plays. Suppose we ignore the constant $A$ and simply solve

$$-a\Delta_g\phi_1 + \phi_1 = |\sigma|^2_g \,.$$

By the maximum principle we have $\phi_1 > 0$, so we can use $\phi_1$ as a conformal factor. Setting $g_1 = \phi_1^{2\kappa}g$ we find that the scalar curvature of of $g_1$ is

$$R_{g_1} = \phi_1^{-2\kappa-1}(-a\Delta_g\phi_1 + \phi_1) = \phi_1^{-2\kappa-1} \, |\sigma|^2_g = \phi_1^{2\kappa+3} \, |\sigma_1|_{g_1}$$

where $\sigma_1 = \phi_1^{-2}\sigma$. Solving the Lichnerowicz equation is then equivalent to solving

$$-a\Delta_{g_1}\phi + R_{g_1}\phi = |\sigma_1|_{g_1} \, \phi^{-2\kappa-3}.$$

Since the zeros of $\sigma_1$ are incorporated into $R_{g_1}$, finding constant sub- and supersolutions is easy. Moreover, there is no obstacle to finding constant sub- and supersolutions even if $\sigma_1$ has singularities.

This motivates our approach for solving (13) when $\lambda_g \geq 0$. We start by making a conformal change to a metric with scalar curvature encoding both the zeros and the singularities of the conformal data, and then we find constant sub- and supersolutions. The specific choice of conformal change replaces specialized arguments used previously for subcases of $\lambda_g \geq 0$ with a single unifying technique.

**Proposition 4.1.** *Suppose* $(M^n, g)$ *is a compact Riemannian manifold with* $g \in H^s$, $s > n/2$ *and* $\lambda_g \geq 0$. *Suppose also that* $\tau$ *is a scalar function in* $H^{s-1}$ *and* $\sigma$ *is a transverse-traceless tensor in* $H^{s-1}$. *Then there exists a positive solution* $\phi \in H^s$ *of the Lichnerowicz equation if and only if one of the following conditions holds:*

1. $\sigma \not\equiv 0$ *and* $\tau \not\equiv 0$.

2. $\lambda_g > 0$, $\sigma \not\equiv 0$ *and* $\tau \equiv 0$.

3. $\lambda_g = 0$, $\sigma \equiv 0$ *and* $\tau \equiv 0$.

*Proof:* We wish to solve

$$-a\Delta_g\phi + R_g\phi = |\sigma|_g^2\,\phi^{-2\kappa-3} - b\tau^2\phi^{2\kappa+1}. \tag{14}$$

From Proposition 3.3 there exists a conformal factor $\phi_1 \in H^s$ such that $g_1 = \phi_1^{2\kappa}g$ has continuous scalar curvature $R_{g_1}$ with the same sign as $\lambda_g$. If $\lambda_g = 0$, $\sigma \equiv 0$ and $\tau \equiv 0$, then $\phi_1$ is a solution of (14) and we are done. So we can restrict our attention to Cases 1 and 2 where $\sigma \not\equiv 0$. Letting $\sigma_1 = \phi_1^{-2}\sigma$, it follows that there exists a solution of (14) if and only if there is a solution of

$$-a\Delta_{g_1}\phi + R_{g_1}\phi = |\sigma_1|_{g_1}^2\,\phi^{-2\kappa-3} - b\tau^2\phi^{2\kappa+1}. \tag{15}$$

From Proposition 6.1, since $\left(R_{g_1} + b\tau^2\right)\phi_2$ is non-negative and not identically zero, there is a solution $\phi_2 \in H^s$ of

$$-a\Delta_{g_1}\phi_2 + \left(R_{g_1} + \frac{2}{3}\tau^2\right)\phi_2 = |\sigma_1|_{g_1}^2$$

Since $\sigma_1 \not\equiv 0$, we have moreover from the maximum principles Lemma 2.9 and Proposition 2.10 that $\phi_2 > 0$. Let $g_2 = \phi_2^{2\kappa}g_1$ and $\sigma_2 = \phi_2^{-2}\sigma_1$. As before, there is a solution of (14) if and only if there is a solution of

$$-a\Delta_{g_2}\phi + R_{g_2}\phi = |\sigma_2|_{g_2}^2\,\phi^{-2\kappa-3} - b\tau^2\phi^{2\kappa+1}. \tag{16}$$

We are now in a setting where we can compute constant sub- and supersolutions. Since

$$R_{g_2} = \left( \left. |\sigma_1| \right._{g_1}^2 - b\tau^2 \phi_2 \right) \phi_2^{-2\kappa-1}$$
$$= |\sigma_2|_{g_2}^2 \phi_2^{2\kappa+3} - b\tau^2 \phi_2^{-2\kappa},$$

a constant $\phi_+$ is a supersolution of (16) if

$$\left[ |\sigma_2|_{g_2}^2 \phi_2^{2\kappa+3} - b\tau^2 \phi_2^{-2\kappa} \right] \phi_+ \geq |\sigma_2|_{g_2}^2 \phi_+^{-2\kappa-3} - b\tau^2 \phi_+^{2\kappa+1}.$$

Pick $\phi_+$ such that $\phi_+^{2\kappa+4} \geq \max_M \phi_2^{-2\kappa-3}$ and such that $\phi_+^{2\kappa} \geq \max_M \phi_2^{-2\kappa}$. This is possible since $\phi_2 \in H^s$ and is hence continuous. Then

$$|\sigma_2|_{g_2}^2 \phi_2^{2\kappa+3} \phi_+ \geq |\sigma_2|_{g_2}^2 \phi_+^{-2\kappa-3}$$
$$-b\tau^2 \phi_2^{-2\kappa} \phi_+ \geq -b\tau^2 \phi_+^{2\kappa+1}$$

and therefore $\phi_+$ is a supersolution. Similarly, pick a positive constant $\phi_-$ such that $\phi_-^{2\kappa+4} \leq \min_M \phi_2^{-2\kappa-3}$ and $0 < \phi_-^{2\kappa} \leq \min_M \phi_2^{-2\kappa}$. Then $\phi_-$ is a subsolution and it is easy to see that $\phi_- \leq \phi+$. So from Propositions 6.2 and 6.4 proved in Section 6 there exists a positive solution $\phi \in H^s$ of (16).

To prove the converse direction, we need to show that if there is a solution with $\lambda_g \geq 0$, and if at least one of $\tau \equiv 0$ or $\sigma \equiv 0$, then either $\lambda_g > 0$ and $\sigma \not\equiv 0$ or $\lambda_g = 0$ and $\sigma$ and $\tau$ both vanish identically. This is a consequence of Proposition 3.3. If a positive solution $\phi$ of (13) exists, we can set $g' = \phi^{2\kappa} g$ and $\sigma' = \phi^{-2}\sigma$ to obtain

$$R_{g'} = |\sigma'|_{g'}^2 - b\tau^2.$$

If $\sigma$ and therefore $\sigma' \equiv 0$, then Corollary 3.4 implies $\lambda_{g'} \leq 0$. Since $\lambda_g \geq 0$ we have $\lambda_g = \lambda_{g'} = 0$ and $\tau^2 \equiv 0$. On the other hand, if $\tau^2 \equiv 0$ and $\sigma \not\equiv 0$, then Corollary 3.4 implies $\lambda_g > 0$. $\square$

Existence in the case $\lambda_g < 0$ is more delicate. Assuming the scalar curvature of $g$ is negative and continuous, we want to follow the approach of Proposition 4.1 and solve an equation such as

$$-a\Delta_g \phi + R_g \phi + b\tau^2 \phi_1 = |\sigma|_g^2 \qquad (17)$$

to make a conformal change to a metric with scalar curvature encoding the zeros and singularities of $\tau$ and $\sigma$. But $R_g < 0$ and we are not assured that we can solve this equation. On the other hand, if $\tau \not\equiv 0$ we can solve

$$-a\Delta_g \phi + b\tau^2 \phi_1 = |\sigma|_g^2$$

as a substitute. Assuming that $\sigma \not\equiv 0$, then $\phi_1$ will be positive and we can set $g_1 = \phi_1^{2\kappa} g$. The resulting scalar curvature

$$R_{g_1} = \phi_1^{-2\kappa-1} \left( |\sigma|_g^2 + R_g \phi_1 - b\tau^2 \phi_1 \right)$$

is problematic. Since $R_g$ is strictly negative the term $R_g - b\tau^2$ does not vanish when $\tau^2$ does. This prevents us from finding constant supersolutions. On the other hand, if $R_g$ were a multiple of $-\tau^2$ then there would be no such difficulty. We now show that the existence of a conformal change to a metric with scalar curvature proportional to $-\tau^2$ is the precise condition required to solve (13) in the negative Yamabe case.

**Proposition 4.2.** *Suppose $(M^n, g)$ is a compact Riemannian manifold with $g \in H^s$, $s > n/2$ and $\lambda_g < 0$. Suppose also that $\tau$ is a function in $H^{s-1}$ and $\sigma$ is a transverse-traceless tensor in $H^{s-1}$. Then there exists a positive solution $\phi \in H^s$ of the Lichnerowicz equation if and only if there is a positive function $\phi_1 \in H^s$ such that $g_1 = \psi_1^{2\kappa} g$ has scalar curvature $R_{\hat{g}_1} = -b\tau^2$.*

*Proof:* In the case $\sigma \equiv 0$ there is nothing to prove we see can assume that $\sigma \not\equiv 0$. First suppose that there exists a conformal factor $\phi_1$ such that $g_1 = \phi_1^{2\kappa} g$ satisfies $R_{g_1} = -b\tau^2$. Since $\lambda_g < 0$ this implies in particular that $\tau \not\equiv 0$. Solving the Lichnerowicz equation then reduces to solving

$$-a\Delta_{g_1}\psi - b\tau^2\psi = |\sigma_1|_{g_1}^2 \, \psi^{-2\kappa-3} - b\tau^2\psi^{2\kappa+1}$$

Let $\phi_2$ be the unique solution in $H^s$ of

$$-a\Delta_{g_1}\phi_2 + b\tau^2\phi_2 = |\sigma_1|_{g_1}^2 \, .$$

Since $\tau^2 \not\equiv 0$ and since $\sigma \not\equiv 0$, the solution exists and is positive. Letting $g_2 = \phi_2^{2\kappa} g_1$ we have

$$R_{g_2} = \phi_2^{-2\kappa-1}(-2b\tau^2\phi_2 + |\sigma_1|_{g_1}^2)$$
$$= -2b\tau^2\phi_2^{-2\kappa} + |\sigma_2|_{g_2}^2 \, \phi_2^{2\kappa+3}.$$

We can now solve

$$-a\Delta_{g_2}\psi + |\sigma_2|_{g_2}^2 \, \phi_2^7\psi - 2b\tau^2\phi_2^{-2\kappa}\psi = |\sigma_2|_{g_2}^2 \, \psi^{-2\kappa-3} - b\tau^2\psi^{2\kappa+1}. \tag{18}$$

by finding constant sub- and supersolutions.

Let $\phi_+$ satisfy $\phi_+^{2\kappa+4} \geq \max_M \phi_2^{-2\kappa-3}$ and $\phi_+^{2\kappa} \geq \max_M 2\phi_2^{-2\kappa}$. Then, arguing as in Proposition 4.1 we have $\phi_+$ is a supersolution. Let $\phi_-$ satisfy $\phi_-^{2\kappa+4} \leq \min_M \phi_2^{-2\kappa-3}$ and $\phi_-^{2\kappa} \leq \min_M 2\phi_2^{-2\kappa}$. Then $\phi_-$ is a subsolution with $\phi_- \leq \phi_+$. Hence by Proposition 6.2 or Proposition 6.4 there exists a solution to (18).

For the converse direction, let us first make a conformal change to a metric with continuous negative scalar curvature via a conformal factor $\phi_1$. Let $g_1 = \phi_1^{2\kappa} g$ and for future reference set $\sigma_1 = \phi_1^{-2}\sigma$. We wish to find a solution of the equation

$$-a\Delta_{g_1}\phi + R_{g_1}\phi = -b\tau^2\phi^{2\kappa+1}. \tag{19}$$

To do this we seek non-constant sub- and supersolutions. Since there exists a positive solution of the Lichnerowicz equation, there also exists a positive solution $\phi_+$ of

$$-a\Delta_{g_1}\phi_+ + R_{g_1}\phi_+ = |\sigma_1|_{g_1}\phi_+^{2\kappa+3} - b\tau^2\phi_+^{2\kappa+1} \geq -b\tau^2\phi_+^{2\kappa+1}.$$

So $\phi_+$ is a supersolution of (19). To find a subsolution we consider the family of equations

$$-a\Delta_{g_1}\phi_\epsilon - R_{g_1}\phi_\epsilon = -R_{g_1} - \epsilon\tau^2.$$

Since $R_{g_1} < 0$ there is a unique solution to this equation for each $\epsilon$. When $\epsilon = 0$ the solution is identically 1. So taking $\epsilon$ sufficiently small, noting that convergence in $H^s$ implies uniform convergence, we can ensure that $\phi_\epsilon > \frac{1}{2}$ everywhere. Fix such an $\epsilon$. We claim that $\phi_- = \eta\phi_\epsilon$ if $\eta$ is chosen sufficiently small. Indeed, pick $\eta$ such that $\eta\phi_\epsilon < \phi_+$ and such that $\eta^{2\kappa} \leq \frac{\epsilon}{b}\min_M \phi_\epsilon^{-2\kappa-1}$. Then

$$\begin{aligned}
-a\Delta_{g_1}\phi_- + R_{g_1}\phi_- &= \eta(R_{g_1}(2\phi_\epsilon - 1) - \epsilon\tau^2)\\
&\leq -\eta\epsilon\tau^2\\
&\leq -b\tau^2(\eta\phi_\epsilon)^{2\kappa+1}\\
&\leq -b\tau^2\phi_-^{2\kappa+1}.
\end{aligned}$$

Hence $\phi_-$ is a subsolution with $0 < \phi_- \leq \phi_+$ and we conclude there exists a positive solution of (19). $\quad\square$

The apparently new observation of Proposition 4.2 is that the study of the Lichnerowicz equation in the Yamabe negative case completely reduces to the study of the prescribed scalar curvature problem. In the smooth category there exists an integral condition [Ra95] we can use to detect if we can make a conformal change to a metric with scalar curvature $-b\tau^2$. Suppose $g$ and $\tau$ are smooth, $\lambda_g < 0$, $R_g$ is a negative constant, and $\mathrm{Vol}_g(M) = 1$ (since we can always make a conformal change to a metric with negative constant scalar curvature and $\mathrm{Vol}_g(M) = 1$, there is no restriction on $g$ here other than $\lambda_g < 0$ and smoothness). Then there exists a conformal change to a metric with scalar curvature $-b\tau^2$ if and only if

$$\inf_{f\in\mathscr{A}} \frac{\int_M a\,|\nabla f|_g^2\,dV_g}{\int f^2\,dV_g} > -R_g \tag{20}$$

where

$$\mathscr{A} = \{f \in H^1 : f \geq 0,\ f \not\equiv 0,\ \text{and}\ \int_M \tau^2 f\,dV_g = 0\}.$$

Since condition (20) is well defined even for rough metrics, it is a good candidate for a necessary and sufficient condition even in the rough category. It remains to be seen, however, if this is true. We prove here a weaker result that suffices for finding CMC solutions of the constraint equations.

**Proposition 4.3.** *Suppose $(M^n, g)$ is a compact Riemannian manifold with $g \in H^s$, $s > n/2$ and $\lambda_g < 0$. Suppose also that $\tau$ a function in $H^{s-1}$ such that $\tau \geq c$ for some positive constant $c$. Then there exists a positive function $\phi \in H^s$ such that $\tilde{g} = \phi^{2\kappa} g$ has scalar curvature $R_{\tilde{g}} = -b\tau^2$.*

*Proof:* We undertake a series of conformal changes designed to build a metric with scalar curvature that is negative, bounded away from zero, and singular wherever $\tau$ is. We first make a conformal change through a function $\phi_1$ given by Proposition 3.3 such that $g_1 = \phi_1^{2\kappa} g$ has continuous negative scalar curvature $R_{g_1}$. Next, let $v \in H^s$ be the solution of

$$-a\Delta_{g_1} v - R_{g_1} v + \tau^2 v = -R_{g_1}$$

which exists and is positive since $R_{g_1} < 0$. Setting $\phi_2 = 1 + v$ we have

$$-a\Delta_{g_1}\phi_2 + R_{g_1}\phi_2 = (2R_{g_1} - \tau^2)v.$$

Hence the scalar curvature of $g_2 = \phi_2^{2\kappa} g_1$ is

$$R_{g_2} = \phi_2^{-2\kappa-1} v(2R_{g_1} - \tau^2).$$

Since $R_{g_1}$, $v$, and $\phi_2$ are all continuous, and since $R_{g_1}$ is negative, we conclude there exist constants $c_1 \geq c_2 > 0$ such that

$$-c_1(1 + \tau^2) \leq R_{g_2} \leq -c_2(1 + \tau^2). \tag{21}$$

To prove the proposition it is enough to find a positive solution of

$$-a\Delta_{g_2}\phi + R_{g_2}\phi = -b\tau^2\phi^5 \tag{22}$$

which we do by finding constant sub- and supersolutions. Pick constants $\phi_-$ and $\phi_+$ such that

$$0 < \phi_-^{2\kappa} < \frac{c_2}{b}$$

and

$$\phi_+^{2\kappa} > \frac{c_1}{b}\left(1 + \frac{1}{\min(\tau^2)}\right).$$

From (21) it follows that $\phi_-$ is a subsolution and $\phi_+$ is a supersolution such that $\phi_- \leq \phi_+$. Hence there exists a positive solution $\phi \in H^s$ of (22). $\quad\square$

The uniqueness of any solution of the Lichnerowicz equation follows, as usual, from the fact that the functions $y \mapsto y^{-2\kappa-3}$ and $y \mapsto -y^{2\kappa}$ are monotone decreasing.

**Proposition 4.4.** *Suppose $(M^n, g)$ is a Riemannian manifold with $g \in H^s$ and $s > n/2$. Suppose also that $\tau$ is a function and $\sigma$ is a transverse-traceless tensor in $H^{s-1}$. If $\phi_1$ and $\phi_2$ are solutions of (13), then either*

1. $\phi_1 = \phi_2$, *or*

2. $\sigma \equiv 0$, $\tau \equiv 0$, $\lambda_g = 0$, *and* $\phi_1 = c\phi_2$ *for some positive constant $c$.*

*Proof:* Let $g_1 = \phi_1^{2\kappa}$, $\sigma_1 = \phi_1^{-2}\sigma$, and $\phi = \phi_2/\phi_1$. Then $\phi$ solves

$$-a\Delta_{g_1}\phi + |\sigma_1|_{g_1}^2 \phi - b\tau^2\phi = |\sigma_1|_{g_1}^2 \phi^{-2\kappa-3} - b\tau^2\phi^{2\kappa+1}.$$

Hence $\phi - 1$ satisfies

$$-a\Delta_{g_1}(\phi - 1) + \left[|\sigma_1|_{g_1}^2 - b\tau^2\right](\phi - 1) = |\sigma_1|_{g_1}^2 (\phi^{-2\kappa-3} - 1) - b\tau^2(\phi^{2\kappa+1} - 1). \quad (23)$$

Since $(\phi - 1)^{(+)} \in H^1$, and since $-a\Delta_{g_1}(\phi - 1) \in H^{s-2} \subset H^{-1}$, we can compute using Lemma 2.8

$$\left\langle -a\Delta_{g_1}(\phi - 1), (\phi - 1)^{(+)} \right\rangle_{(M,g_1)} = a \left\langle \nabla(\phi - 1), \nabla(\phi - 1)^{(+)} \right\rangle_{(M,g_1)}$$
$$= \int_{\phi > 1} a \left\langle \nabla(\phi - 1), \nabla(\phi - 1) \right\rangle_{g_1} dV_g.$$

Hence

$$\int_{\phi > 1} a \left\langle \nabla(\phi - 1), \nabla(\phi - 1) \right\rangle_{g_1} dV_g = $$
$$\int_{\phi > 1} |\sigma_1|_{g_1}^2 (\phi^{-2\kappa-3} - \phi)(\phi - 1) + b\tau^2(\phi - \phi^{2\kappa+1})(\phi - 1) dV_g \quad (24)$$

If $\phi > 1$ then $(\phi^{-2\kappa-3} - \phi)(\phi - 1) < 0$ and $(\phi - \phi^{2\kappa+1})(\phi - 1) < 0$. So the integral on the right hand side of (24) must be non-positive. Since the integral on the left-hand side is non-negative, we conclude

$$\int_{\phi > 1} a \left\langle \nabla(\phi - 1), \nabla(\phi - 1) \right\rangle_{g_1} dV_{g_1} = 0 \quad (25)$$

$$\int_{\phi > 1} |\sigma_1|_{g_1}^2 (\phi^{-2\kappa-3} - \phi)(\phi - 1) + b\tau^2(\phi - \phi^{2\kappa+1})(\phi - 1) dV_{g_1} = 0. \quad (26)$$

A similar argument using $(\phi - 1)^{(-)}$ as a test function shows

$$\int_{\phi < 1} a \left\langle \nabla(\phi - 1), \nabla(\phi - 1) \right\rangle_{g_1} dV_{g_1} = 0 \quad (27)$$

$$\int_{\phi < 1} |\sigma_1|_{g_1}^2 (\phi^{-2\kappa-3} - \phi)(\phi - 1) + b\tau^2(\phi - \phi^{2\kappa+1})(\phi - 1) dV_{g_1} = 0. \quad (28)$$

From the continuity of $\phi - 1$ and equations (25) and (27) we conclude $\phi$ is constant and therefore $\phi_1$ is a constant multiple of $\phi_2$. Suppose that $\phi_1 \neq \phi_2$, so $\phi \neq 1$. Then from equation (26) or (28) we obtain

$$\int_M |\sigma_1|_{g_1}^2 (\phi^{-2\kappa-3} - \phi)(\phi - 1) + b\tau^2(\phi - \phi^{2\kappa+1})(\phi - 1) dV_{g_1} = 0$$

and therefore $|\sigma_1|_{g_1}^2 \equiv 0$, $\tau^2 \equiv 0$ and $\lambda_g = 0$. $\quad \square$

We can now establish in the rough setting the necessary and sufficient conditions under which the Lichnerowicz equation is solvable when $\tau$ is constant.

**Theorem 4.5.** *Suppose $(M^n, g)$ is a compact Riemannian manifold with $g \in H^s$ and $s > n/2$. Suppose also that $\tau$ is a constant and $\sigma$ is a transverse-traceless tensor in $H^{s-1}$. Then there exists a positive solution $\phi \in H^s$ of the Lichnerowicz equation (13) if and only if one of the following conditions holds:*

1. *$\sigma \not\equiv 0$ and $\tau \neq 0$.*

2. *$\lambda_g > 0$, $\sigma \not\equiv 0$ and $\tau = 0$.*

3. *$\lambda_g = 0$, $\sigma \equiv 0$ and $\tau = 0$.*

4. *$\lambda_g < 0$, $\sigma \equiv 0$ and $\tau \neq 0$.*

*In Cases 1, 2, and 4 the solution is unique. In Case 3 the solution is unique up to multiplication by a positive constant.*

*Proof:* The proof of the proposition in the case $\lambda_g \geq 0$ follows directly from Propositions 4.1 and 4.4. The existence of a solution if $\lambda_g < 0$ and $\tau \neq 0$ follows from Propositions 4.2 and 4.3; its uniqueness is a consequence of Proposition 4.4. So it only remains to show that if $\lambda_g < 0$ and if there is a solution, then $\tau \neq 0$. But this is a consequence of Corollary 3.4, arguing as at the end of Proposition 4.1.  □

**Remark.** It is easy to extend these results to so-called scaled matter sources. In the presence of scaled sources, the Lichnerowicz equation becomes

$$-a\Delta_g \phi + R_g \phi = |\sigma|_g^2 \phi^{-2\kappa-3} + \rho\phi^{-1-\kappa} - b\tau\phi^{\kappa+1} \tag{29}$$

where $\rho \in H^{s-2}$ and $\rho \geq 0$. The previous theorems carry over to this case by replacing the condition $\sigma \equiv 0$ (resp. $\sigma \not\equiv 0$) with $\sigma \equiv 0$ and $\rho \equiv 0$ (resp. $\sigma \not\equiv 0$ or $\rho \not\equiv 0$). For example, the initial step of Proposition 4.2 when $\lambda_g \geq 0$ and $\sigma \not\equiv 0$ or $\rho \not\equiv 0$ is to solve

$$-a\Delta_g \phi + b\tau^2\phi = |\sigma|_g^2 + \rho$$

rather than

$$-a\Delta_g \phi + b\tau^2\phi = |\sigma|_g^2 .$$

We have only used features of the term $|\sigma|_g^2 \phi^{-2\kappa-3}$ that are shared by the term $\rho\phi^{-\kappa 1}$.

# 5. The Momentum Constraint

The treatment of the momentum constraint for rough metrics follows easily from the a priori estimates of Corollary 2.6 and standard arguments. We have the following properties of the vector Laplacian $\Delta_{\mathbb{L}} = \operatorname{div} \mathbb{L}$.

**Proposition 5.1.** *Suppose $(M^n, g)$ is a compact Riemannian manifold with $g \in H^s$ and $s > n/2$. Then $\Delta_{\mathbb{L}} : H^s \to H^{s-2}$ is Fredholm of index 0. The kernel of $\Delta_{\mathbb{L}}$ consists of conformal Killing fields in $H^s$ and the equation*

$$\Delta_{\mathbb{L}} X = Y$$

*with $Y \in H^{s-2}$ is solvable if and only if*

$$\langle Y, Z \rangle_{(M,g)} = 0$$

*for every conformal Killing field $Z \in H^s(M)$.*

*Proof:* The estimates of Corollary 2.6 imply that $\Delta_{\mathbb{L}}$ is semi-Fredholm. On the other hand, the vector Laplacian of a smooth metric is well known to have index 0. Approximating $g$ with a sequence of smooth metrics, using the fact that the index of a semi-Fredholm map is locally constant, it follows that the index of $\Delta_{\mathbb{L}}$ is also 0. If $\Delta_{\mathbb{L}} X = 0$, then integration by parts yields $0 = -\langle \Delta_{\mathbb{L}} X, X \rangle_{(M,g)} = \langle \mathbb{L} X, \mathbb{L} X \rangle_{(M,g)}$ so $X$ is a conformal Killing field. Finally, the conditions under which the equation $\Delta_{\mathbb{L}} X = Y$ is solvable follow from the fact that $\Delta_{\mathbb{L}}$ is Fredholm and self-adjoint. $\square$

The York decomposition of symmetric, trace-free, (0,2)-tensors is a standard consequence of Proposition 5.1.

**Theorem 5.2.** *Suppose $(M^n, g)$ is a Riemannian manifold of with $g \in H^s$ and $s > n/2$. Let $\mathscr{S}$ be the space of symmetric trace-free (0,2) tensors in $H^{s-1}$. Then*

$$\mathscr{S} = \operatorname{im} \mathbb{L} \oplus \ker \operatorname{div}$$

*where $\mathbb{L}$ acts on $H^s$ vector fields and $\operatorname{div}$ acts on $\mathscr{S}$.*

*Proof:* Suppose $\sigma \in \mathscr{S}$. If $X$ is a conformal Killing field, then $\langle \operatorname{div} \sigma, X \rangle_{(M,g)} = -\langle \sigma, \mathbb{L} X \rangle_{(M,g)} = 0$. So from Proposition 5.1 there exists a vector field $W \in H^s$ such that $\Delta_{\mathbb{L}} W = \operatorname{div} \sigma$. Since $\sigma = \mathbb{L} W + (\sigma - \mathbb{L} W)$, and since $\operatorname{div}(\sigma - \mathbb{L} W) = 0$, we have proved that $\mathscr{S} = \operatorname{im} \mathbb{L} + \ker \operatorname{div}$. To show that the sum is a direct sum, it suffices to show that $\operatorname{im} \mathbb{L} \cap \ker \operatorname{div}$ is trivial. If $\sigma \in \operatorname{im} \mathbb{L} \cap \ker \operatorname{div}$, then $\sigma = \mathbb{L} W$ for some $W \in H^s$. Since $\operatorname{div} \sigma = 0$, we have $\Delta_{\mathbb{L}} W = 0$ and hence $W$ is a conformal Killing field. So $\sigma = \mathbb{L} W = 0$. $\square$

# 6. Sub- and Supersolutions

We now turn to the existence theorem for the semilinear problem

$$-\Delta_g u = F(x, u) \tag{30}$$

where $F$ has the form

$$F(x, y) = \sum_{j=1}^{l} F_j(x) G_j(y).$$

We start with the following properties of the linearized operator.

**Proposition 6.1.** *Suppose $(M^n, g)$ is a compact Riemannian manifold with $g \in H^s$ and $s > n/2$. Suppose also that $V \in H^{s-2}$ and let $\mathscr{L} = -\Delta_g + V$. Then $\mathscr{L} : H^\sigma \to H^{\sigma-2}$ for any $\sigma \in [2-s, s]$ is Fredholm with index 0 and we have the estimate*

$$||u||_{H^\sigma} \lesssim ||\mathscr{L}u||_{H^{\sigma-2}} + ||u||_{H^{2-s}}. \tag{31}$$

*The kernel of $\mathscr{L}$ is a finite dimensional subspace of $H^s$. If $V \geq 0$ and $V \not\equiv 0$, then the kernel of $\mathscr{L}$ is trivial and we have the estimate*

$$||u||_{H^\sigma} \lesssim ||\mathscr{L}u||_{H^{\sigma-2}}. \tag{32}$$

*If $f \in H^{\sigma-2}$, the equation*

$$\mathscr{L}u = f$$

*admits a solution if and only if $\langle f, v \rangle_{(M,g)} = v$ for every $v \in \ker \mathscr{L}$. Finally, if $u \in H^\sigma$ for some $\sigma \in [2-s, s]$, and if $\mathscr{L}u \in H^{\tau-2}$ for some $\tau \in [\sigma, s]$, then $u \in H^\tau$.*

*Proof:* First consider the Laplacian $\Delta_g$. The estimates of Corollary 2.6 imply that $\Delta_g$ is semi-Fredholm acting on $H^\sigma$ with $\sigma \in (2-s, s]$. On the other hand, the Laplacian of a smooth metric is well known to have index 0. Approximating $g$ with a sequence of smooth metrics, using the fact that the index of a semi-Fredholm map is locally constant, it follows that the index $\Delta_g$ is also 0. With the help of Lemma 2.1 it is easy to see that $\mathscr{L}$ is a compact perturbation of $-\Delta_g$ and we conclude $\mathscr{L}$ acting on $H^\sigma$ with $\sigma \in (2-s, s]$ is is also Fredholm with index 0. Since $\mathscr{L}$ is self-adjoint, it follows that $\mathscr{L}$ is in fact Fredholm acting on $H^\sigma$ with $\sigma \in [2-s, s]$. The estimate (31) is trivial if $\sigma = 2-s$ and follows from Corollary 2.6 for $\sigma \in (2-s]$. Since the index of $\mathscr{L}$ is 0, and since $\mathscr{L}$ is self adjoint, the dimension of the kernel of $\mathscr{L}$ acting on $H^s$ is the same as the dimension of the kernel of $\mathscr{L}$ acting on $H^{2-s}$. So these kernels coincide, which shows that the kernel of $\mathscr{L}$ acting on $H^\sigma$ for any $\sigma \in [2-s, s]$ is contained in $H^s$; the kernel is finite dimensional since $\mathscr{L}$ is Fredholm. If $V \geq 0$ and $V \not\equiv 0$, then the weak maximum principle Lemma 2.9 implies the kernel of $\mathscr{L}$ is trivial and estimate (32) follows since $\mathscr{L}$ is invertible. The conditions under which the equation

$$\mathscr{L}u = f$$

is solvable now follow from standard Fredholm theory. Finally, suppose $u \in H^\sigma$ with $\sigma \in [2-s, s]$ and suppose

$$\mathscr{L}u = f$$

with $f \in H^{\tau-2}$ where $\sigma \leq \tau \leq s$. We want to show that $u \in H^\tau$. To see this, we note that since $\langle f, v \rangle_{(M,g)} = \langle \mathscr{L}u, v \rangle_{(M,g)} = 0$ for every $v \in \ker \mathscr{L}$ we can solve

$$\mathscr{L}\hat{u} = f$$

for some $\hat{u} \in H^\tau$ and therefore $u - \hat{u} \in \ker \mathscr{L} \subset H^s$. Hence $u \in H^\tau$.   $\square$

The existence theorem for the equation

$$-\Delta_g u = \sum_{j=1}^{l} F_j(x) G_j(u). \tag{33}$$

is a technical refinement of the constructive approach used in many previous papers treating the constraint equations [CB04][CBIY00][Is95]. In [Ma04] the technique was extended to asymptotically Euclidean manifolds $(M^n, g)$ of class $H^s$ with $s > n/2$. That proof divided into two cases: 3-manifolds and higher dimensional manifolds. We follow the same approach here and start with a result for the hard case of 3-manifolds that treats a more general class of coefficient functions $F_j$ than those considered in [Ma04].

**Proposition 6.2.** *Suppose*

1. *$(M^3, g)$ is a compact Riemannian manifold with $g \in H^s$ and $s > 3/2$,*

2. *$u_-, \ u_+ \in H^s$ are a subsolution and a supersolution respectively of (30) such that $u_- \leq u_+$,*

3. *each $F_j \in H^{s-2}(M)$ and is non-negative,*

4. *each $G_j$ is smooth on $I = [\inf(u_-), \sup(u_+)]$.*

*Then there exists a solution $u \in H^s(M)$ of (30) such that $u_- \leq u \leq u_+$.*

*Proof:*   We first assume $3/2 < s \leq 2$. Let

$$V(x) = 1 + \sum_{j=1}^{l} F_j(x) \left| \min_I G_j' \right|$$

so that $V \in H^{s-2}(M)$ is nonnegative and not identically zero. Let $F_V(x, y) = F(x, y) + V(x)y$ and so that $F_V$ is non-decreasing in $y$. By this we mean that if $u, v \in H^s$ satisfy $u \geq v$, then $F_V(x, u) - F_V(x, v) \geq 0$ as a distribution. Finally, let $L_V = -\Delta_g + V$ so that $L_V : H^s \to H^{s-2}$ is an isomorphism.

We construct a sequence of functions as follows. Let $u_0 = u_+$, and for $i \geq 1$ let $u_i$ be the solution of

$$L_V \, u_{i+1} = F_V(x, u_i).$$

Now

$$L_V(u_1 - u_0) \leq F_V(x, u_0) - F(x, u_0) - V(x)u_0 = 0$$

since $u_0$ is a super-solution. Moreover,

$$L_V(u_1 - u_-) \geq F_V(x, u_+) - FV(x, u_-) \geq 0$$

since $u_+ \geq u_-$. From Lemma 2.9 we conclude $u_0 \geq u_1 \geq u_-$. Now suppose $u_0 \geq u_1 \geq \cdots \geq u_i \geq u_-$. Then

$$L_V(u_{i+1} - u_i) = F_V(x, u_i) - F(x, u_{i-1}) \leq 0.$$

Hence $u_{i+1} \leq u_i$. Also,

$$L_V(u_{i+1} - u_-) = F_V(x, u_i) - F(x, u_-) \geq 0.$$

So $u_i \geq u_{i+1} \geq u_-$. We obtain by induction for the entire sequence $u_+ = u_0 \geq u_1 \geq u_2 \geq \cdots \geq u_-$.

We claim the sequence $\{u_i\}_{i=1}^{\infty}$ is bounded in $H^s(M)$. From Proposition 6.1 we can estimate

$$||u_{i+1}||_{H^s(M)} \lesssim ||F_V(x, u_i)||_{H^{s-2}(M)}. \tag{34}$$

Now $F_V(x, u_i)$ is a sum of terms of the form

$$F(x)G(u_i)$$

where $F \in H^{s-2}(M)$. Fix $s' \in (3/2, s)$. From Lemma 6.3 proved below we have

$$||F(x)G(u_i)||_{H^{s-2}} \lesssim ||F||_{H^{s-2}} \big[ ||G(u_i)||_{L^\infty} + ||G'(u_i)||_{L^\infty} ||u_i||_{H^{s'}} \big].$$

Since $u_- \leq u_i \leq u_+$, we have uniform estimates for each of the terms $||G(u_i)||_{L^\infty}$ and $||G'(u_i)||_{L^\infty}$. Hence

$$||F_V(x, u_i)||_{H^{s-2}(M)} \lesssim 1 + ||u||_{H^{s'(M)}}. \tag{35}$$

Now from interpolation we know that for any $\epsilon > 0$ there is a constant $C(\epsilon)$ such that

$$||u_i||_{H^{s'}} \leq C(\epsilon)||u_i||_{L^2} + \epsilon||u_i||_{H^s}. \tag{36}$$

Again, since $u_- \leq u_i \leq u_+$, we have uniform estimates for $||u_i||_{L^2}$. Combining (35) and (36), taking $\epsilon$ sufficiently small, we obtain

$$||u_{i+1}||_{H^s(M)} \leq \frac{1}{2}||u_i||_{H^s(M)} + C$$

for some constant $C$ independent of $i$. Iterating this inequality we obtain a bound for all $i$

$$||u_i||_{H^s} \leq ||u_+||_{H^s} + 2C.$$

Hence some subsequence of $\{u_i\}_{i=1}^{\infty}$ (and by monotonicity, the whole sequence) converges weakly in $H^s(M)$ to a limit $u_{\infty}$.

It remains to see $u_{\infty}$ is a solution of (30). Now $u_i$ converges strongly to $u_{\infty}$ in $H^{s'}$ for any $s' < s$ and also converges uniformly on compact sets. Hence for any $\phi \in C^{\infty}(M)$,

$$\int_M \left( F_V(x, u_i) - V(x)u_{i+1} \right) \phi \, dV_g \to \int_M F(x, u_{\infty}) \phi \, dV_g$$

$$\int_M \langle \nabla u_{i+1}, \nabla \phi \rangle \, \phi \, dV_g \to \int_M \langle \nabla u_{\infty}, \nabla \phi \rangle \, dV_g.$$

So

$$\int_M \langle \nabla u_{\infty}, \nabla \phi \rangle \, dV_g = \int_M F(x, u_{\infty}) \phi \, dV_g,$$

which shows that $u_{\infty}$ is a weak solution. So $-\Delta_g u_{\infty} = F(x, u_{\infty})$ in the sense of distributions.

To handle the case $s > 2$ we use a bootstrap. First suppose $4 \geq s \geq 2$. From the above we have a solution $u$ in $H^2$. Since $2 > n/2 = 3/2$ and since $2 > s - 2 \in [0, 2]$, we know from Lemma 2.2 that $c(x)f(u) \in H^{s-2}$. Since $-\Delta_g u \in H^{s-2}$, Proposition 6.1 implies $u \in H^s$. We now obtain the result for all $s > 3/2$ by iteration.  □

The following technical lemma completes the proof of Proposition 6.2.

**Lemma 6.3.** *Suppose $(M^n, g)$ is a Riemannian manifold with $g \in H^s$ and $s > n/2$. Suppose also that $F$ is a smooth function, $\sigma \in [-1, 1]$, and $s' \in (n/2, s)$. Then for every $u \in H^s$ and $v \in H^{\sigma}$ we have*

$$||vf(u)||_{H^{\sigma}(M)} \lesssim ||v||_{H^{\sigma}} \left[ ||f(u)||_{L^{\infty}(M)} + ||f'(u)||_{L^{\infty}(M)}||u||_{H^{s'}(M)} \right].$$

*Proof:* If $\sigma = 1$, then

$$||vf(u)||_{H^1(M)} \lesssim ||vf(u)||_{L^2(M)} + ||\nabla(vf(u))||_{L^2(M)}$$

$$\lesssim ||vf(u)||_{L^2(M)} + ||\nabla vf(u)||_{L^2(M)} + ||vf'(u)\nabla u||_{L^2(M)}$$

$$\lesssim ||v||_{H^1(M)}||f(u)||_{L^{\infty}(M)} + ||vf'(u)\nabla u||_{L^2_{(}M)}.$$

To estimate the last term we note that since $v \in H^1$, it is also in $L^p$ where

$$\frac{1}{p} = \frac{1}{2} - \frac{1}{n}.$$

On the other hand, $\nabla u \in L^q$ where

$$\frac{1}{q} = \frac{1}{2} - \frac{s' - 1}{n}.$$

Since

$$\begin{aligned}
\frac{1}{p} + \frac{1}{q} &= \frac{1}{2} - \frac{1}{n} + \frac{1}{2} - \frac{s' - 1}{n} \\
&= \frac{1}{2} + \left( \frac{1}{2} - \frac{s'}{n} \right) \\
&< \frac{1}{2}
\end{aligned}$$

we conclude

$$||vf'(u)\nabla u||_{L^2(M)} \lesssim ||v||_{H^1} ||f'(u)||_{L^\infty(M)} ||u||_{H^{s'}}. \tag{37}$$

This proves the result in the case $s = 1$.

We prove the case $s = -1$ by duality. If $w \in H^1(M)$, then

$$\begin{aligned}
\left| \langle vf(u), w \rangle_{(M,g)} \right| &\lesssim ||v||_{H^{-1}} ||f(u)w||_{H^1(M)} \\
&\lesssim ||v||_{H^{-1}} \Big[ ||f(u)||_{L^\infty(M)} + ||f'(u)||_{L^\infty(M)} ||u||_{H^{s'}(M)} \Big] ||w||_{H^1(M)}.
\end{aligned}$$

Hence

$$||vf(u)||_{H^{-s}} \lesssim ||v||_{H^{-1}} \Big[ ||f(u)||_{L^\infty(M)} + ||f'(u)||_{L^\infty(M)} ||u||_{H^{s'}(M)} \Big].$$

We have therefore obtained the result for $s = -1$, and the result for all $s \in [-1, 1]$ now follows from interpolation. □

The proof of Proposition 6.2 does not apply to manifolds $M^n$ with $n > 3$ since it depends, for example, on the fact that $H^2$ is an algebra. From the point of view of the constraint equations, this is not a serious drawback since the most interesting dimension is $n = 3$. Nevertheless, it would be nice to have an existence theorem in higher dimensions. This can be done following the approach of [Ma04]. The idea is to first obtain existence in the space $W^{k,p}$ where $k$ is the largest integer with $k < n$ and where $\frac{1}{p} = \frac{1}{2} - \frac{s-k}{n}$. The full regularity in $H^s$ then follows from an interpolation argument and elliptic regularity. The proof from [Ma04] readily adapts to the compact setting using results and techniques from [CB04] in place of [Ma03] to obtain existence in $W^{k,p}$. The bootstrap argument then follows exactly as in [Ma04]; we omit the proof.

**Proposition 6.4.** *Suppose*

1. $(M^n, g)$ *is a Riemannian manifold with* $n > 3$, $g \in H^s$ *and* $s > n/2$,

2. $u_-$, $u_+ \in H^s$ *are a subsolution and a supersolution respectively of (30) such that* $u_- \leq u_+$,

3. *each* $F_j \in H^{s-2}(M)$ *and is non-negative,*

4. *each* $G_j$ *is smooth on* $I = [\inf(u_-), \sup(u_+)]$.

*Then there exists a solution* $u \in H^s(M)$ *of (30) such that* $u_- \leq u \leq u_+$.

**Acknowledgements.** This work was sponsored in part by NSF grant DMS-0305048 and the University of Washington Royalty Research Fund.